\documentclass[useAMS,usenatbib]{mn2e}

\usepackage{graphics}

\title[Towards a dynamical mass of NGC~5408~X-1]{Towards a dynamical mass of the ultraluminous X-ray source NGC~5408~X-1}
\author[D. Cseh et al.]{D. Cseh$^{1,5}$\thanks{E-mail:
d.cseh@astro.ru.nl}, F. Gris\'{e}$^{2,3,4}$, P. Kaaret$^{4}$, S. Corbel$^{5}$, S. Scaringi$^{6,1}$,  P. Groot$^{1}$, H. Falcke$^{1,7,8}$ 
 \newauthor and E. K\"{o}rding$^{1}$ 
\\
$^{1}$Department of Astrophysics/IMAPP, Radboud University Nijmegen, P.O. Box 9010, 6500 GL Nijmegen, The Netherlands\\
$^{2}$Instituto de Astrof\'isica de Canarias, E-38200 La Laguna, Tenerife, Spain\\
$^{3}$Departamento de Astrof\'isica, Universidad de La Laguna, Avda. Astrof\'isico Francisco Sanchez s/n, E-38271 La Laguna, Tenerife, Spain\\
$^{4}$Department of Physics and Astronomy, Iowa City, 52240, USA\\
$^{5}$Laboratoire Astrophysique des Interactions Multi-echelles (UMR 7158),\\ 
CEA/DSM-CNRS-Universite Paris Diderot, CEA Saclay, F-91191 Gif sur Yvette, France\\
$^{6}$Instituut voor Sterrenkunde, KU Leuven, Celestijnenlaan 200D, B-3001 Leuven, Belgium\\
$^{7}$ASTRON, Dwingeloo, The Netherlands\\
$^{8}$Max-Planck-Institut f\"{u}r Radioastronomie Bonn, Germany
 }

\begin{document}

\date{Draft}

\pagerange{\pageref{firstpage}--\pageref{lastpage}} \pubyear{2013}

\maketitle

\label{firstpage}

\begin{abstract}

We obtained multi-epoch Very Large Telescope (VLT) optical spectroscopic data in 2011 and 2012 on the ultraluminous X-ray source (ULX) NGC~5408~X-1. We confirm that the He {\sc ii} $\lambda$4686 line has a broad component with an average FWHM of $v= 780\pm64$ km/s with a variation of $\sim$13\% during observations spanning over 4 years, and is consistent with the origin in the accretion disc. The deepest optical spectrum does not reveal any absorption line from a donor star. Our aim was to measure the radial velocity curve and estimate the parameters of the binary system. We find an upper limit on the semi-amplitude of the radial velocity of $K=132\pm42$ km/s. A search for a periodic signal in the data resulted in no statistically significant period. The mass function and constraints on the binary system imply a black hole mass of less than $\sim 510$~M$_{\odot}$. Whilst, a disc irradiation model may imply a black hole mass smaller than $\sim 431-1985$~M$_{\odot}$, depending on inclination. Our data can also be consistent with an unexplored orbital period range from a couple of hours to a few days, thus with a stellar-mass black hole and a subgiant companion.

\end{abstract}


\section{Introduction}

Ultraluminous X-ray sources (ULXs) are non-nuclear, accreting binary systems in external galaxies with an X-ray luminosity greater than the Eddington-limit of a 20-M$_{\odot}$ black hole ($L_X > 3\times10^{39}$ erg/s), assuming isotropic emission. Several explanations exist for the ULX phenomenon: these sources could possibly be mechanically or relativistically beamed, possibly accrete above their Eddington-limit, or represent the high end of the population of stellar-mass black holes with masses of 20-100 M$_{\odot}$ \citep[see review][]{Feng:2011kx}. A class of ULXs, called Hyperluminous X-ray sources (HLXs), are the best candidate intermediate-mass black holes (IMBHs; $10^{2}-10^{5}$ M$_{\odot}$) due to their high X-ray luminosities of $L_X > 10^{41}$ erg/s \citep{Kaaret:2006ys,Farrell:2009ys,Walton:2011uq,Sutton:2012kx,Webb:2012kx}.   

The remedy for the degeneracies between the abovementioned theoretical models is dynamical mass measurements of the binary system. However, it is difficult to obtain any radial velocity curve due to the fact that the most nearby ULXs lie at a distance of 3-5 Mpc, which results in faint ($V\sim 22-24$ mag) optical counterparts. Furthermore, many optical counterparts show variable emission indicative of a dominant contribution from an accretion disc rather than a donor star \citep{Tao:2011kx,Grise:2012li,Soria:2012fk,Gladstone:2013fk}. So far only two ULXs: P13 \citep{Motch:2011fk} and M101-ULX1 \citep{Liu:2009uq} show absorption lines from a late type B supergiant and emission lines from a Wolf-Rayet donor, respectively. Their typical X-ray luminosity is $\sim$10$^{39}$ erg/s. 

On the other hand, one can study the periodic variation of the X-ray \citep[e.g.][]{Kaaret:2006ys,Foster:2013fk} or the optical \citep[e.g.][]{Zampieri:2012kx} light curve of ULXs, or the broad permitted lines in the optical spectra. Broad permitted lines can be produced in the dense regions of the accretion disc reflecting the velocity distribution of the inner disc. The velocity shifts of such lines may trace the orbital motion of the system and provide constraints on the mass of the compact object \citep{Hutchings:1987vn}. In principle, the He {\sc ii} $\lambda$4686 line is the most promising tracer due to its highest ionization potential in the optical band. However, several studies were made using this method, e.g. on Holmberg~IX~X-1, NGC~1313~X-2 \citep{Roberts:2011bh,Liu:2012fk}, NGC~5408~X-1 \citep{Cseh:2011hc} resulting in no firm mass constraints. This can be due to several reasons, like insufficient time sampling of the radial velocity curve, insufficient spectral resolution resulting in blurred narrow and broad components, differing viewing geometries of the system, or a non-disc origin of the He {\sc ii} line. We also note that in a number of low-mass X-ray binaries (LMXBs) and cataclysmic variables (CVs) the radial velocity variation of disc emission lines mostly traces the orbital period, but the associated amplitude is likely to be off from the true amplitude \citep[e.g.][]{Cantrell:2007fk,Groot:2001uq}.  

NGC~5408~X-1 is a bona fide ULX reaching an X-ray luminosity of $10^{40}$ erg/s. It is surrounded by a photoionized optical nebula \citep{Kaaret:2009uq,Grise:2012li,Cseh:2012fk} and it is one of the few ULXs associated with a radio nebula \citep{Kaaret:2003pz,Lang:2007qz,Cseh:2012fk}. The energetics of the optical nebula suggests that this ULX is, at most, mildly beamed \citep{Kaaret:2009uq} leaving the possibility of a stellar-mass black hole accreting at super-Eddington rates or a sub-Eddington but more massive stellar-mass black hole.

\citet{Strohmayer:2009cr} reported an orbital period of $\sim 115$ day inferred from the X-ray light curve. On the other hand \citet{Foster:2010qy} suggested that this period is superorbital in nature. Four years of Swift/XRT monitoring revealed that after a few cycles this period disappeared and therefore is unlikely to be the orbital period \citep{Grise:2013vn}. However, quasi-periodic dipping behaviour with a period of $250\pm28$ day was revealed \citep{Grise:2013vn,Pasham:2013fk}. \citet{Pasham:2013fk} argue that the dipping period may reflect the orbital period of the system or can occur due to a precessing accretion disc as well.

The optical counterpart of the ULX point source was identified by \citet{Lang:2007qz} and its X-ray to NIR SED was studied by \citet{Grise:2012li}, who found that it can be consistent with either a disc irradiation model or a B0I supergiant. The optical spectrum of NGC~5408~X-1 shows permitted and forbidden emission lines indicative of an X-ray photoionized nebula, and no absorption lines from a donor star. A higher resolution optical spectrum revealed the presence of broad and narrow components of permitted lines indicative of a distinct origin: likely the disc and the nebula, respectively \citep{Cseh:2011hc}. The He {\sc ii} line is markedly shifted between two observations and assuming that this shift traces the orbital motion of the system an upper limit  on the mass of the black hole of $\sim 1800$ M$_{\odot}$ was set \citep{Cseh:2011hc}.

Here we present results of multi-epoch VLT spectroscopic observations from 2011 and 2012 that aimed to measure the radial velocity curve of NGC~5408~X-1. We describe our observations in Sec. 2., our results in Sec 3., and discuss the binary parameters in Sec. 4.

\section{Observations}

We conducted multi-epoch VLT observations in order to measure the radial velocity curve of NGC~5408~X-1 (PI: Cseh). FORS-2 observations were obtained in 2011 and in 2012 using the 1200B grism with a slit width of $0.7\arcsec$ covering the spectral range of 3660$-$5110 \AA\ with a dispersion of 0.36 \AA\ pixel$^{-1}$ and a spectral resolution of $\lambda/\Delta\lambda=1420$ at the central wavelengths. Each observing block (OB) consisted of three 850~s exposures in 2011 and three 880~s exposures in 2012 with a 12 pixel offset along the spatial axis between successive exposures. We note that OBs from 2008 were taken with a lower resolution 600B grism. See Table \ref{obs} for more details. We note that OB9 was re-observed due to seeing conditions, however the data quality was good enough to include in our analysis. On the other hand, only 11 out of 20 OBs were scheduled and observed in 2012 due to weather conditions. CCD pixels were binned by 2 in both the spatial and spectral dimensions. 

\begin{table}
\begin{center}
\caption{VLT FORS observations of NGC~5408~X-1\label{obs}}
\begin{tabular}{lccc}
\hline
OB & Start  & Duration [s]& Seeing ["]\\
\hline
OBa&2008-04-08 04:59:04& 3 $\times$ 850&0.78\\
OBb&2008-04-08 05:46:04& 3 $\times$ 850&0.80\\
OBc&2008-04-08 06:45:13& 3 $\times$ 850&0.89\\
OBd&2008-04-09 05:46:17& 3 $\times$ 850&1.14\\
OBe&2008-04-10 05:08:33& 3 $\times$ 850&0.67\\
OBf&2008-04-10 05:54:29& 3 $\times$ 850&0.67\\
\hline
OBg&2010-04-12 07:46:55& 3 $\times$ 850&0.46\\
\hline
OB1&2011-04-03 06:31:33& 3 $\times$ 850&0.66\\
OB2&2011-04-11 05:23:49& 3 $\times$ 850&0.63\\
OB3&2011-04-27 05:55:08& 3 $\times$ 850&0.74\\
OB4&2011-05-09 04:33:47& 3 $\times$ 850&0.64\\
OB5&2011-05-23 02:27:10& 3 $\times$ 850&0.98\\
OB6&2011-05-26 01:26:51& 3 $\times$ 850&0.87\\
OB7&2011-06-09 05:21:40& 3 $\times$ 850&0.65\\
OB8&2011-06-26 03:15:13& 3 $\times$ 850&0.66\\
OB9a&2011-06-29 00:36:08& 3 $\times$ 850&1.15\\
OB9b&2011-07-01 02:17:41& 3 $\times$ 850&0.64\\
OB10&2011-07-02 02:07:09&1 $\times$ 418 + 3 $\times$ 850&0.63\\
\hline
OB11&2012-04-15 02:55:28& 3 $\times$ 880&0.65\\
OB12&2012-04-22 05:59:24& 3 $\times$ 880&0.48\\
OB13&2012-04-23 03:55:59& 3 $\times$ 880&0.66\\
OB14&2012-05-18 04:03:11& 3 $\times$ 880&0.70\\
OB15&2012-05-20 05:16:41& 3 $\times$ 880&0.64\\
OB16&2012-06-18 01:08:53& 3 $\times$ 880&0.79\\
OB17&2012-06-21 00:25:03& 3 $\times$ 880&1.02\\
OB18&2012-06-23 01:26:44& 3 $\times$ 880&0.74\\
OB19&2012-07-18 23:49:35& 3 $\times$ 880&0.86\\
OB20&2012-08-12 23:54:35& 3 $\times$ 880&1.00\\
OB21&2012-08-14 23:47:47& 3 $\times$ 880&0.79\\

\end{tabular}
\end{center}
\medskip The table shows the start date of each OBs in UT, the duration of successive exposures and the averaged seeing corrected by airmass during each observation. OBa--OBf are adapted from \citet{Kaaret:2009uq}, OBg is adapted from \citet{Cseh:2011hc}.
\end{table}

We reduced the data using the Image Reduction and Analysis Facility (IRAF)\footnote{IRAF is distributed by the National Optical Astronomy Observatory, which is operated by the Association of Universities for Research in Astronomy, Inc., under cooperative agreement with the National Science Foundation.} \citep{Tody:1993fk}. After correcting the three exposures in each OB using bias and flat-field images, we aligned and median averaged them to eliminate bad pixels and cosmic rays using the {\tt imcombine} task with the {\tt ccdclip} rejection algorithm.

Due to the dimness of the ULX counterpart we used a bright nearby star at 2MASS position $\alpha_{\rm J2000} = 14^h 03^m 18.^s97, \delta_{\rm J2000} = -41\degr 22\arcmin 56.\arcsec6$ as a reference trace. The slit position angle was --105.71$^{\circ}$ (N=0$^{\circ}$, E=90$^{\circ}$). The trace for the ULX counterpart was centered on the He {\sc ii} $\lambda 4686$ emission line profile. For the He {\sc ii} line profile analysis, we used a trace width of 4 pixels corresponding to $1.0\arcsec$ in order to isolate the ULX emission from the nebular emission. The HgCdHeNeAr lamp and standard stars provided by ESO were used for wavelength and flux calibration. An atmospheric extinction correction was applied using the IRAF built-in Cerro Tololo Inter-American Observatory (CTIO) extinction tables. We corrected for the reddening using $E(B-V)=0.08 \pm 0.03$ \citep{Kaaret:2009uq,Cseh:2011hc} and the extinction curve from \citet{Cardelli:1989uq} with $R_{V}=3.1$.

\section{Analysis and Results}

Following \citet{Cseh:2011hc}, in each OB we fitted the He {\sc ii} line with a two-component Gaussian profile. First, we fitted the continuum with a second-order polynomial around each line, excluding the line itself by visual examination. The flux errors were estimated by the root mean square deviation of the data in the same region. For the two-component Gaussian fit we used a non-linear least squares fit by employing the {\tt LMFIT} subroutine, that is based on "MRQMIN" \citep{Press:1992qf}, of the Interactive Data Language (IDL) version 7.0. 

\begin{figure}
\resizebox{3.35in}{!}{
\includegraphics{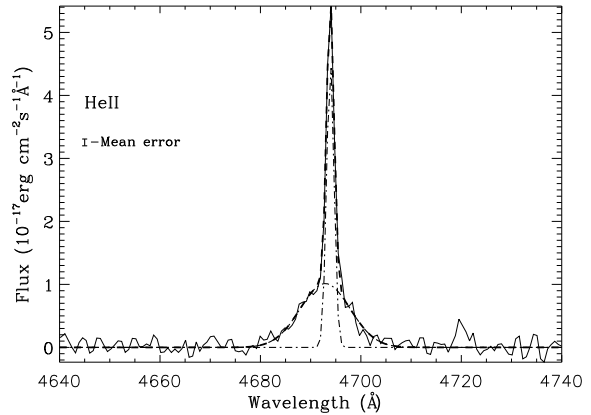}}
\caption{The figure shows the two-component Gaussian fit of the He {\sc ii} line of OB6 as a demonstration. The separate components are shown with dashed-dotted lines, the overall fit is shown as a dashed line.}
\label{ob6}
\end{figure}

We obtained a good fit in all but five OBs; we summarise our results in Table \ref{broad} and show OB6 as an example in Fig \ref{ob6}. As noted by \citet{Cseh:2011hc}, in the 2008 data we did not detect a broad component in OBa, OBb, OBd. Similarly, in our current data set, OB1, OB3, OB7, OB20 and OB21 do not show any broad component, which is probably due to variations in its flux (see later in Sec. 4.).

\begin{table*}
\begin{center}
\caption{The broad component of the He {\sc ii} line \label{broad}}
\begin{tabular}{lcccccc}
\hline
OB&MJD  & Wavelength & $v_s$ & $v_{s,err}$ & FWHM &Flux \\
& &  [\AA] & [km/s] & [km/s] & [\AA] &\\
\hline
OBc&54564.291740&	4692.08 $\pm$	0.77&148.5 & 49.3&11.09 $\pm$ 1.44&0.56 $\pm$ 0.20\\
OBe&54566.224608&	4690.03 $\pm$	1.43&16.5 & 91.3&11.75 $\pm$ 1.33&0.49 $\pm$ 0.13\\
OBf&54566.256508&	4689.94 $\pm$	1.70&10.8 & 108.8&10.82 $\pm$ 1.55&0.53 $\pm$ 0.15\\
\hline
OBg&55298.334603	&	4694.63	$\pm$	0.25	&310.2 & 16.0&	12.37	$\pm$	0.69	&	1.12	$\pm$	0.08\\
\hline
OB2&55662.235202	&	4692.04	$\pm$	0.26	&145.1 & 16.6&	10.51	$\pm$	0.62	&	0.81	$\pm$	0.05\\
OB4&55690.200478	&	4693.99	$\pm$	0.24	&257.5 & 15.4&	13.89	$\pm$	0.66	&	0.85	$\pm$	0.04\\
OB5&55704.112531	&	4692.36	$\pm$	0.26	&147.1 & 16.6&	13.01	$\pm$	0.63	&	0.85	$\pm$	0.04\\
OB6&55707.070653	&	4692.88	$\pm$	0.22	&179.1 & 14.1&	11.69	$\pm$	0.57	&	1.01	$\pm$	0.05\\
OB8&55738.145904	&	4693.75	$\pm$	0.11	&223.7 & 7.0&	10.42	$\pm$	0.31	&	1.69	$\pm$	0.06\\
OB9a&55741.035417	&	4693.02	$\pm$	0.27	&176.2 & 17.3&	12.53	$\pm$	0.71	&	0.80	$\pm$	0.05\\
OB9b&55743.105957	&	4694.68	$\pm$	0.20	&281.9 & 12.8&	12.86	$\pm$	0.55	&	1.26	$\pm$	0.06\\
OB10&55744.095981	&	4692.12	$\pm$	0.27	&117.8 & 17.3&	13.64	$\pm$	0.64	&	1.98	$\pm$	0.10\\
\hline
OB11&56032.132541	&	4690.56	$\pm$	0.79	&48.4 & 50.5&	14.28	$\pm$	1.75	&	0.31	$\pm$	0.04\\
OB12&56039.260256	&	4694.11	$\pm$	0.46	&272.4 & 29.4&	14.19	$\pm$	1.24	&	0.49	$\pm$	0.04\\
OB13&56040.174559	&	4691.39	$\pm$	0.76	&98.0 & 48.6&	16.31	$\pm$	1.81	&	0.40	$\pm$	0.04\\
OB14&56065.179567	&	4693.42	$\pm$	0.48	&216.7 & 30.7&	10.69	$\pm$	1.33	&	0.38	$\pm$	0.05\\
OB15&56067.230624	&	4691.08	$\pm$	0.76	&66.1 & 48.6&	12.17	$\pm$	1.59	&	0.30	$\pm$	0.04\\
OB16&56096.058511	&	4692.20	$\pm$	0.27	&126.8 & 14.1&	10.56	$\pm$	0.63	&	0.70	$\pm$	0.05\\
OB17&56099.028092	&	4691.02	$\pm$	0.41	&50.3 & 26.2&	10.33	$\pm$	0.80	&	0.60	$\pm$	0.05\\
OB18&56101.070920	&	4692.48	$\pm$	0.35	&143.1 & 22.4&	11.55	$\pm$	0.83	&	0.60	$\pm$	0.05\\
OB19&56127.003432	&	4693.90	$\pm$	0.61	&228.4 & 39.0&	11.90         $\pm$	1.65	&	0.32	$\pm$	0.05\\
\end{tabular}
\end{center}
\medskip The table shows the result of the Gaussian fit for the broad component of the He {\sc ii} line. The OB number, the wavelength, the full-width half maximum and the flux is presented. The flux is in units of $10^{-17}$ erg cm$^{-2}$ s$^{-1}$ \AA$^{-1}$. OBc, OBe, OBf, OBg are adapted from Table 1. in \citet{Cseh:2011hc}. MJD values are corresponding to the middle of each OB. The table also shows the barycentric corrected velocity shift ($v_s$) with respect to the reference point. The error on the velocity shift ($v_{s,err}$) corresponds to the wavelength uncertainty in Col. 3. 
\end{table*}

\subsection{The broad component}

The average FWHM of the broad component is $W_{b}=12.2\pm1.0$ \AA \,with a standard deviation of 1.6 \AA. Given that the standard deviation is greater than the error on $W_b$, we estimate the width variation to be $\sim$13\%. The line profiles are Gaussian and the corresponding average FWHM in velocity units is $v= 780\pm64$ km/s. The rms wavelength shift of the broad component of $2.89\pm0.65$ \AA, corresponding to $185\pm42$ km/s, is a substantial fraction of the line width\footnote{ We note that the co-added spectrum (Fig. 6.) shows an apparent broad component of the H$\beta$ line confirming previous results of \citet{Cseh:2011hc}, however, most of the OBs are not suitable for similar analysis due to a relatively low flux in the broad component.}.  

To place a conservative upper limit on the semi-amplitude of the radial velocity excursion, we quote half of the difference between the maximum and minimum wavelength shift of the broad component (excluding OBs from 2008 that have high uncertainties due to a different spectral resolution). We find an upper limit  on the semi-amplitude of $132\pm42$ km/s.

As a next step, we tested whether the wavelength shift of the broad component can be due to motion within the accretion disc, e.g. an outflowing material may cause a blue shift and a narrower line width and vice versa. To this purpose we performed correlation tests, between the wavelength ($\lambda$) and the FWHM of the broad component, between the flux of the broad component and $\lambda$, and between the flux and the FWHM. We find no correlation in any of these parameters (Table \ref{cor}).

\begin{table}
\begin{center}
\caption{Correlation tests\label{cor}}
\begin{tabular}{lcc}
\hline
Test & Kendall's $\tau$  & Probability\\
\hline
$\lambda$ vs FWHM&0.085&     0.62 \\
$\lambda$ vs Flux&0.29 &0.086\\
FWHM vs Flux &-0.072&0.67\\
\end{tabular}
\end{center}
\medskip The table shows correlation tests between the fitted parameters of the broad component of the He {\sc ii} line.
\end{table}

\subsection{Period Search}

As a next step, we performed a period search using three different methods: first, we used an epoch folding method, then a phase dispersion minimisation (PDM), and finally a Lomb-Scargle (LS) periodogram analysis.
 
For the epoch folding period search, first, we obtained helio- and barycentric velocity corrections using an IDL implemented code called {\tt baryvel} that is accurate to velocities of $\sim$1 m/s and based on the algorithm of \citet{Stumpff:1980fk}. Given that the zero point of the velocity curve is unknown, we set a reference frame to the bluest shift of the broad component; see Table \ref{broad}. This results in a velocity curve that is shifted by a constant value which we fit with an additional free parameter ($v_0$). We note that setting the reference frame to the narrow component of the He {\sc ii} line may introduce bias on the shift of the broad component. This is because a nebular component can have a complex velocity structure as well as variations of $\sim$50-80 km/s \citep{Lehmann:2005bu}. 

\begin{figure}
\resizebox{3.5in}{!}{
\includegraphics{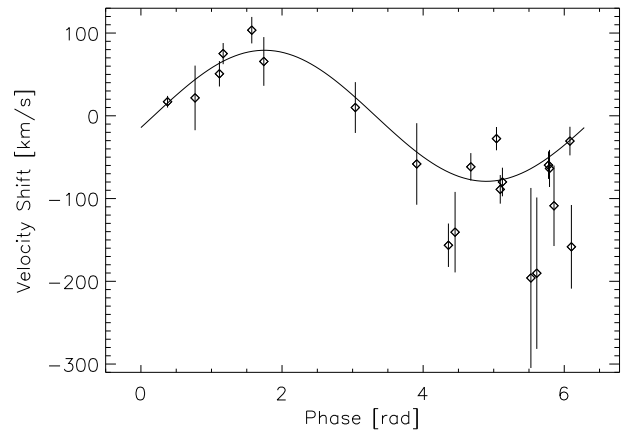}}
\caption{The figure shows an example fit of the velocity curve folded on the best fit test period.}
\label{rv}
\end{figure}

Recent X-ray studies of NGC~5408~X-1 reveal a quasi-periodic dipping behaviour of $\sim$250$\pm$28 day \citep{Grise:2013vn,Pasham:2013fk}. Motivated by these results, we folded over a period range from 1 day up to 300 day with a resolution of $10^{-4}$ day to match the uncertainty on the epoch of the observations. Using {\tt LMFIT}, described above, each of these folded velocity curves were then fitted with a sinusoidal function: $v_r(\Phi)=K \cos(\Phi-\Phi_0)+v_0$, where $K$ is the amplitude in units of km/s, $\Phi$ is the phase, $\Phi_0$ is the phase shift in units of radian, $v_0$ is a constant in units of km/s. We calculated the phase for each epoch as $\Phi=2\pi (1 - \frac{t\, {\rm mod}\, P}{P})$, where $t$ is the MJD date and $P$ is the test period. We find a best fit test period of $P=2.6464$ day for a $\chi^2$/d.o.f=2.8 with a semi-amplitude of the velocity curve of $K=83\pm6$ km/s, and $v_0=212\pm$6 km/s, and $\Phi_0=1.55\pm0.11$ in units of radian, see Fig. \ref{rv}. 

\begin{figure}
\resizebox{3.5in}{!}{
\includegraphics{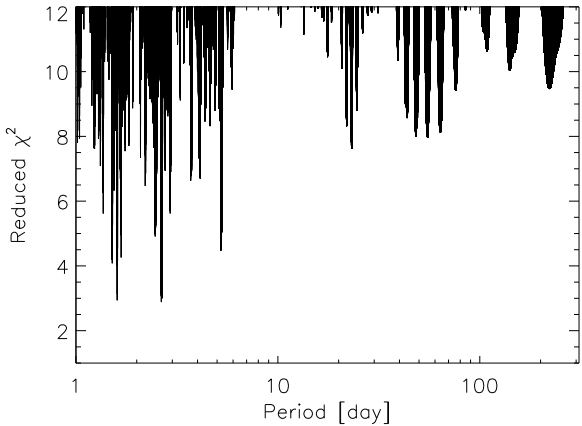}}
\caption{The figure shows the reduced $\chi^{2}$ values of the sinusoidal fit as a function of period. For clarity, $\chi^{2}$ values below 12 are plotted.}
\label{redc}
\end{figure}

In order to assess the quality of the fit we evaluated the corresponding reduced $\chi^{2}$ values of each trial period (see Fig. \ref{redc}). Using the same fitting method as above, we also evaluated the $\chi^{2}$ value of a constant and a linear fit for a comparison. The constant fit resulted in a reduced $\chi^{2}$ of 13.5 that is independent of the period range. On the other hand, for the linear fit we used a period range identical to the range of the sinusoidal fit, and minimising the obtained $\chi^{2}$ values resulted in $\chi^{2}/d.o.f.= 2.9$. A comparison to the minimum $\chi^{2}$ of the sinusoidal fit shows that there is no real period found (see below as well). 

\begin{figure}
\resizebox{3.35in}{!}{
\includegraphics{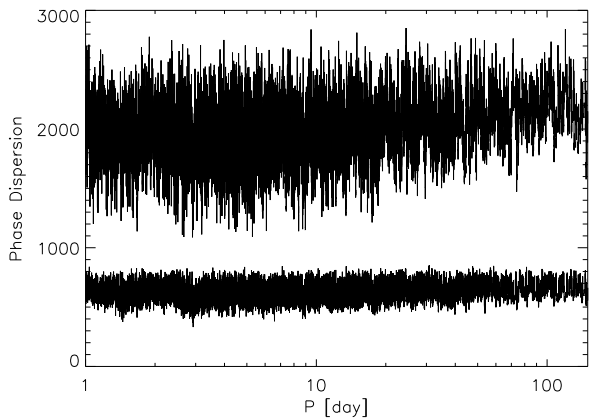}}
\caption{Phase dispersion diagram of the velocity shift of the broad component in NGC~5408~X-1. The bottom values are corresponding to a 90\% significance.}
\label{pdm}
\end{figure}

We also searched for nonsinusoidal periods using a string-length method \citep{Dworetsky:1983fk}, which is a binless phase dispersion minimisation \citep{Stellingwerf:1978fk} suitable for sparse, unevenly sampled measurements. We folded our velocity curves and calculated phases as described above. Then, we evaluated the sum of the squared differences of the folded, phase ordered, velocity curves corresponding to each test period.  On the other hand, we also estimated the significance for a period range of 1 to 150 days using 10$^4$ white noise simulations. We computed 90\% probabilities for each period step of 10$^{-4}$ day separately and found no statistically significant period (Fig. \ref{pdm}).

\begin{figure}
\resizebox{3.35in}{!}{
\includegraphics{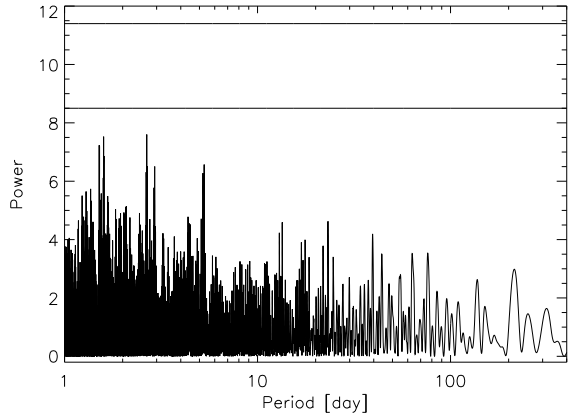}}
\caption{Lomb-Scargle periodogram of the velocity shift of the broad component in NGC~5408~X-1. The horizontal lines correspond to 99\% and 90\% significance, respectively.}
\label{ls}
\end{figure}

Finally, we performed Lomb-Scargle periodogram analysis \citep{Scargle:1982fk} using the {\tt scargle}\footnote{http://astro.uni-tuebingen.de/software/idl/aitlib/timing/\\scargle.html} IDL subroutine. We estimated the significance of the periods using white noise simulations and found that there is no period above a 90\% probability (Fig. \ref{ls}). 

In summary, the data do not show any statistically significant period, regardless of the methods we use, either due to the sparse sampling or due to other effects (see Sec. 4.).

\subsection{Donor star?}

\begin{figure}
\resizebox{3.25in}{!}{
\includegraphics{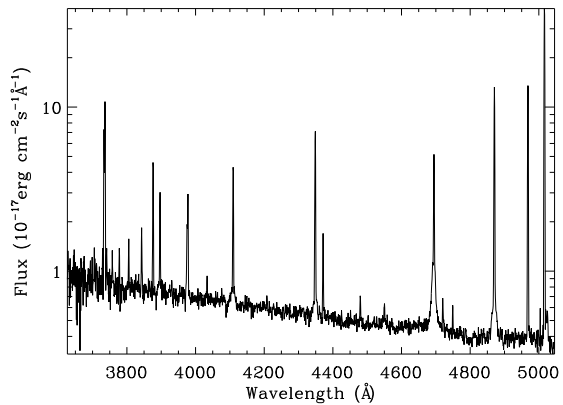}}
\caption{The figure shows the co-added spectrum of ULX NGC~5408~X-1 on a log scale. No redshift correction was applied.}
\label{avg}
\end{figure}

In the following we describe our search for absorption lines that could be a signature of the donor star. To increase our signal to noise ratio in the optical spectrum, we averaged together all the OBs from 2010 to 2012, see Fig. \ref{avg}. We note that we did not include OBs from 2008 due to a different spectral resolution. 

First, we searched for possible nebular lines \citep{Acker:1989uq} and identified previously unknown emission lines. We identified further transitions of He {\sc i} and He {\sc ii} and also an argon line, see Table \ref{lines}. To further test the nature of the emission lines, we also averaged the OBs using a wider trace of 16 pixels -- i.e. allowing more flux from the nebula -- to compare the flux of the lines to the data with the narrow trace. We find that the flux of the emission lines are increased and do not show any width variation (see Table \ref{lines}), therefore the newly identified emission lines likely originate from the nebula.

Then, we made a systematic search on stellar lines using an online catalog\footnote{http://cdsarc.u-strasbg.fr/viz-bin/nph-Cat/html?VI\%2F26\%2F}. In particular, we looked for all highly ionized metallic absorption lines like Si, O, C, Mg that are typical for a B star. We find no absorption features in our optical spectrum. We find a typical rms level of $2.6\times10^{-19}$ erg cm$^{-2}$ s$^{-1} \AA^{-1}$ increasing to $3.8\times10^{-19}$ erg cm$^{-2}$ s$^{-1} \AA^{-1}$ in a range of 5000-4000 \AA. 

\begin{table}
\begin{center}
\caption{Optical lines from NGC~5408~X-1\label{lines}}
\begin{tabular}{lccc}
\hline
Name & Wavelength [\AA]  & Flux & FWHM [\AA] \\
\hline
[Ar {\sc iv}] 4740 & 4750.0 &4.3 $\pm$ 0.3&1.6 $\pm$ 0.2\\
&&9.7 $\pm$ 0.3& 1.7 $\pm$ 0.2\\
\hline
He {\sc i} 4711 & 4720.8 &5.3 $\pm$ 0.5&2.1 $\pm$ 0.2\\
&&12 $\pm$ 1 & 1.9 $\pm$ 0.2\\
\hline
He {\sc ii} 4541 & 4550.8 &5.0 $\pm$ 0.7 &3.0 $\pm$ 0.4\\
&&9.0 $\pm$ 0.5& 3.0 $\pm$ 0.3\\
\hline
He {\sc i} 4471 & 4480.7 &5.9 $\pm$ 0.4&2.2 $\pm$ 0.2\\
&&19 $\pm$ 1&2.0 $\pm$ 0.1\\
\hline
He {\sc i} 4026 & 4034.5 &4.3 $\pm$ 0.4&1.3 $\pm$ 0.2\\
&&8.5 $\pm$ 0.5&1.4 $\pm$ 0.1\\
\end{tabular}
\end{center}
\medskip The table shows emission lines identified in addition to Table 2. in \citet{Kaaret:2009uq}. The flux is in units of $10^{-18}$ erg cm$^{-2}$ s$^{-1}$. The values in the second lines correspond to a wider trace of 16 pixel.  
\end{table}

\section{Discussion}

In the following section we attempt to characterise and constrain some of the physical parameters of the binary system. First, we only use the information on the width of the broad component, then we discuss the radial velocity curve, and finally consider the effects of disc irradiation.  

We recall that the average FWHM of the broad component is $W_{b}=12.2\pm1.0$ \AA \,with a width variation of $\sim$13\%. This is much less than in the case of NGC~1313~X-2 that shows FWHM variation between 3-10 \AA \citep{Grise:2009fk,Roberts:2011bh}. Also, the line width in NGC~5408~X-1 remains relatively stable during the different epochs of observations, that span 4 years. The average FWHM of $v= 780\pm64$ km/s is consistent with production in an accretion disc. Furthermore, the small variability may indicate a relatively stable line emitting region. Assuming that the gas has a Keplerian motion around the black hole and the disc fills $\sim$70\% of the Roche-lobe radius, we can set limits on the system parameters using Eq 4. of  \citet{Groot:2001fk} that we parametrized for NGC~5408~X-1:

\begin{equation}
M_{BH}= 1.89 \times 10^{-8} \left(\frac{v}{\rm{km/s}}\right)^{3} \left(\frac{P}{\rm{day}}\right) \sin^{-3}i\,\,\, \rm{M_{\odot}}.
\end{equation}

Given that the X-ray light curve of NGC~5408~X-1 shows quasi-periodic dipping behaviour, which is reminiscent to those of the high-inclination LMXBs \citep{Grise:2013vn}, the accretion disc is probably not face-on. In order to set a lower limit on the mass of the black hole ($M_{BH}$), we assume an edge-on system ($i=90^{\circ}$). This results in $M_{BH,min}= 9 \left(\frac{P}{\rm{day}}\right)$~M$_{\odot}$. On the other hand, taking a minimum inclination of $\sim 20^{\circ}$, would result in $M_{BH}= 224 \left(\frac{P}{\rm{day}}\right)$~M$_{\odot}$. Using these constraints on $M_{BH}$, the minimum orbital period of the binary system would be in the range of 19 min to 8 hr. Small periods, in the order of hours \footnote{To evaluate the validity of a $\sim$19 min period, we repeated our analysis on each 15-min long individual exposure of OB8. OB8 was selected because it is moderately affected by cosmic rays around the He {\sc ii} line and it has a high signal to noise ratio of the broad component allowing us to fit these individual exposures. We find no evidence of wavelength shift (or FWHM change) between these individual exposures.}, are not ruled out by current observations as significant wavelength shifts are detected between consecutive observations obtained within 24~h. On the other hand, based on the hydrogen rich optical spectrum, periods below an hour are likely to be ruled out even if the smallest main sequence donor is considered \citep[e.g.][]{Verbunt:1995fk}.

Furthermore, one can estimate the inner radius of the line-emitting region ($R$), that corresponds to a Keplerian motion of the gas as assumed above \citep{Porter:2010ve,Groot:2001fk}; we find that $R=2.18 \times 10^{11} \sin^{2} i \left(\frac{M_{BH}}{\rm{10\,\,\, M_{\odot}} }\right)$~cm, which we utilise in the following sections.

\subsection{Radial velocity curve}

Given that no orbital period was found, we set a conservative upper limit on the semi-amplitude of the radial velocity of $K=132\pm42$ km/s ( Sec. 3.1). It was empirically found that if $K$ is inferred from disc emission line shift, then it may also be smaller by a factor of up to 2.5 than the true value \citep{Cantrell:2007fk,Liu:2012fk}. Additionally, \citet{Grise:2013vn} argue that a precessing tilted disc or a warped accretion disc could also be responsible for the dipping behaviour and the highly variable X-ray count rate during these dipping periods. So, the wavelength shift could be biased by variable accretion rate and by partial illumination of the disc that could contribute to the apparent centroid shifts. Moreover, it was shown for some well studied binary systems, that the outer part of the disc could be heavily affected or even dominated by the hot spot region, leading to distorted or nonsinusoidal radial velocity curves \citep{Groot:2001uq}.

\begin{figure}
\resizebox{3.2in}{!}{
\includegraphics{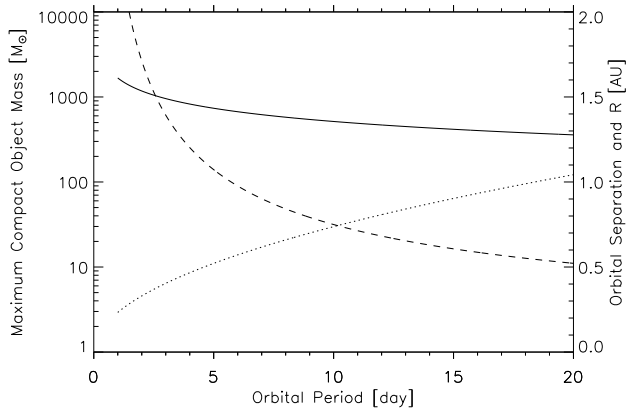}}
\caption{The figure shows the binary parameters as a function of orbital period for a specific set input of parameters (see text). The solid line shows the mass of the black hole (left scale). The orbital separation is shown as dotted line and the dashed line shows the size of the line-emitting region (both with values on the right scale).}
\label{binper}
\end{figure}

On the other hand, the rms wavelength shift of the broad component of $185\pm42$ km/s is a substantial fraction of the line width. The line width is relatively stable and there is no correlation between e.g. the width and the shift of the broad component of the He {\sc ii} line (see Sec. 3.2). Assuming that the shift is a fraction of the true radial velocity excursion, one can characterise the mass of the black hole as a function of period. We combine the mass function and the argument that the radius of the line-emitting region ($R$) is smaller than the orbital separation ($a$). First, we obtain the mass of the black hole as a function of period using Eq. 2 of \citet{Cseh:2011hc}:
\begin{equation}\label{bin}
M_{BH}=(M_c \sin i)^{\frac{3}{2}} \left(\frac{PK^3}{2\pi G}\right)^{-\frac{1}{2}} - M_c  
\end{equation}
where $M_c$ is the mass of the companion, $G$ is the gravitational constant. To obtain an upper limit one has to assume a maximum companion mass, an edge-on system (possibly consistent with the dipping behaviour), and a minimum semi-amplitude. A typical companion mass of $M_c=20$~M$_{\odot}$ is suggested by the extensive stellar environment studies of ULXs, including NGC~5408~X-1 \citep{Grise:2008ly,Grise:2011uq,Grise:2012li}. Albeit a minimum semi-amplitude on the radial velocity curve is not known, we set its value to $K=30$ km/s, which is a factor of $\sim$5 lower than our conservative upper limit and is in the range of the usually considered minimum values (see simulations by \citet{Liu:2012fk}). As a next step, one can calculate the radius of the line-emitting region as a function of period using the obtained black hole mass from Eq.~\ref{bin}. Similarly, one can evaluate the orbital separation as function of period taking into account $M_c$. The argument of $a>R$ limits the period and consequently the mass of the black hole (Fig. \ref{binper}). For these bounds, we obtain a mass of the black hole of less than $\sim 510$~M$_{\odot}$.

\subsection{Disc irradiation and disc size}

Disc irradiation or super-Eddington accretion can modify the disc geometry and opacity, thus the amount and the radius where the X-ray (or total) luminosity is thermalised. However, these models might be mutually exclusive because, in the case of super-Eddington accretion, the soft X-rays are interpreted as due to a wind/outflow, that would not show reprocessed optical emission \citep{Poutanen:2008fk,Kaaret:2009uq}. Additionally, super-Eddington sources may show beamed emission \citep{Middleton:2013uq}. 

This is in contrast to NGC~5408~X-1, that does not show significant beaming \citep{Kaaret:2009uq}. Its SED can be well fitted with a disc irradiation model, resulting in an outer disc radius of $R_{out}=(1-5) \times 10^{12}$~cm for a face-on disc \citep{Grise:2012li}. Also, we recall that the X-ray count rate drops by a factor of 35 or more below average on a quasi-periodic basis \citep{Grise:2013vn}. In addition, the broad component of the He {\sc ii} line was not visible during several observations, spread across years (in 2008 OBa, OBb, OBd, in 2011 OB1, OB3, OB7, in 2012 OB20 and OB21). Except OB7, all of these observations were obtained outside the X-ray coverage. OB7 falls into a time range where dips might be expected (see Fig. 1 and Fig 3. in \citet{Grise:2013vn}). Considering the large width of the broad component as well, the optical emission is likely due to disc irradiation. 

So, assuming that disc irradiation is the main contributor to the optical emission, one may compare the outer disc radius ($R_{out}$) to the inner radius of the line-emitting region ($R$, see before Sec. 4.1.) to constrain the mass of the black hole. \citet{Copperwheat:2007fk} argue that the relative contribution of disc irradiation with respect to irradiation by a donor star scales as $\cos i$, i.e. no contribution from the disc for an edge-on system and vice versa. To have a dominant disc contribution one has to consider $\cos i > 0.5$, i.e. $i<60^{\circ}$. Comparing the radii, $R$ and $R_{out}$, and taking into account that $R_{out} \propto \cos^{-0.5} i$, we find a black hole mass of $\sim 431-1985$~M$_{\odot}$ for $60^{\circ} > i > 20^{\circ}$. On the other hand, it is uncertain whether the region where the He~{\sc ii} line genesis takes place extends to the outer disc radius derived from the disc irradiation model. Therefore, this mass estimate should be taken as an upper limit.

\section{Conclusion}

We studied the optical spectrum of the ultraluminous X-ray source NGC~5408~X-1 using multi-epoch VLT observations in 2011 and in 2012.  We attempted to measure the radial velocity curve using the broad component of the He {\sc ii} emission line, that is consistent with an origin in the accretion disc. The data do not show any statistically significant periodicity and we obtained an upper limit on the semi-amplitude of the putative radial velocity curve of $K=132\pm42$ km/s. Thanks to the number of observations obtained, we identified previously unknown nebular lines; and we found no absorption lines that could be indicative of a donor star.

We also showed that the width of the broad component is relatively stable and its flux is variable, which we interpret as an effect of disc irradiation. A disc irradiation model would imply a black hole mass smaller than $\sim 431-1985$~M$_{\odot}$, depending on inclination. Using a different approach and a conservative set of binary parameters, we find that the mass of the black hole is less than $\sim 510$~M$_{\odot}$. Our results do not rule out the presence of a stellar mass black hole, and suggest that future optical observations should probe an unexplored period range from a couple of hours to a few days.

\section*{Acknowledgments}

Based on observations made with ESO Telescopes 
at the La Silla or Paranal Observatories under program ID 087.D-0156(A) and 089.D-0646(A).

\bibliographystyle{mn2e} 

\bibliography{my}

\label{lastpage}

\end{document}